\documentclass[11pt,a4paper,twocolumn]{article}

\usepackage{graphics}
\usepackage{amssymb}

\newcommand{\captionfonts}{\footnotesize} 

\makeatletter  
\long\def\@makecaption#1#2{%
  \vskip\abovecaptionskip
  \sbox\@tempboxa{{\captionfonts #1: #2}}%
  \ifdim \wd\@tempboxa >\hsize
    {\captionfonts #1: #2\par}
  \else
    \hbox to\hsize{\hfil\box\@tempboxa\hfil}%
  \fi
  \vskip\belowcaptionskip}
\makeatother   

\long\def\symbolfootnote[#1]#2{\begingroup%
\def\thefootnote{\fnsymbol{footnote}}\footnote[#1]{#2}\endgroup}

\begin{document}
\onecolumn
\begin{center}
\LARGE 
A detection system to measure muon-induced neutrons for direct Dark Matter searches\symbolfootnote[3]{\it Accepted for publication in Astroparticle Physics}
\end{center}

\vspace{5mm}

\begin{center}
\small
The EDELWEISS collaboration,\\
V.Yu.~Kozlov\symbolfootnote[1]{Corresponding author, Valentin.Kozlov@kit.edu}$^{,\;a)}$,
E.~Armengaud$\;^{b)}$,
C.~Augier$\;^{c)}$,
A.~Benoit$\;^{d)}$,
L.~Berg\'e$\;^{e)}$,
O.~Besida$\;^{b)}$,
J.~Bl\"umer$\;^{a),\;f)}$,
A.~Broniatowski$\;^{e)}$,
V.~Brudanin$\;^{g)}$,
A.~Chantelauze$\;^{a)}$, 
M.~Chapellier$\;^{e)}$,
G.~Chardin$\;^{e)}$,
F.~Charlieux$\;^{c)}$,
S.~Collin$\;^{e)}$,
X.~Defay$\;^{e)}$, 
M.~De~J\'esus$\;^{c)}$,
P.~Di~Stefano\footnote{Present address: Department of Physics, Queen's University, Kingston, ON, K7L 3N6, Canada}$^{,\;c)}$,
Y.~Dolgorouki$\;^{e)}$,
J.~Domange$\;^{b),\;e)}$,
L.~Dumoulin$\;^{e)}$,
K.~Eitel$\;^{a)}$, 
J.~Gascon$\;^{c)}$,
G.~Gerbier$\;^{b)}$,
M.~Gros$\;^{b)}$,
M.~Hannawald$\;^{b)}$,
S.~Herv\'e$\;^{b)}$,
A.~Juillard$\;^{c)}$,
H. Kluck$\;^{a)}$, 
R.~Lemrani$\;^{b)}$,
P.~Loaiza$\;^{i)}$,
A.~Lubashevskiy$\;^{g)}$,
S.~Marnieros$\;^{e)}$, 
X.-F.~Navick$\;^{b)}$,
E.~Olivieri$\;^{e)}$,
P.~Pari$\;^{h)}$,
B.~Paul$\;^{b)}$,
S.~Rozov$\;^{g)}$,
V.~Sanglard$\;^{c)}$,
S.~Scorza$\;^{c)}$,
S.~Semikh$\;^{g)}$, 
A.S.~Torrent\'o-Coello$\;^{b)}$,
L.~Vagneron$\;^{c)}$,
M-A.~Verdier$\;^{c)}$,
E.~Yakushev$\;^{g)}$
\end{center}

\vspace{4pt}

\begin{center}
\footnotesize \it 
\noindent $^{a)}${Karlsruhe Institute of Technology, Institut f\"ur Kernphysik, Postfach 3640, 76021 Karlsruhe, Germany}\\
\noindent $^{b)}${CEA, Centre d'\'Etudes Nucl\'eaires de Saclay, IRFU, 91191 Gif-sur-Yvette Cedex, France}\\
\noindent $^{c)}${Institut de Physique Nucl\'eaire de Lyon, Universit\'e de Lyon (Universit\'e Claude Bernard Lyon 1) et IN2P3-CNRS, 4 rue Enrico Fermi, 69622 Villeurbanne, France}\\
\noindent $^{d)}${Institut N\'eel, CNRS, 25 Avenue des Martyrs, 38042 Grenoble cedex 9, France}\\
\noindent $^{e)}${Centre de Spectroscopie Nucl\'eaire et de Spectroscopie de Masse, UMR8609 IN2P3-CNRS, Univ. Paris Sud, b\^at 108, 91405 Orsay Campus, France}\\
\noindent $^{f)}${Karlsruhe Institute of Technology, Institut f\"ur Experimentelle Kernphysik, Gaedestr. 1, 76128 Karlsruhe, Germany}\\
\noindent $^{g)}${Laboratory of Nuclear Problems, JINR, Joliot-Curie 6, 141980 Dubna, Moscow Region, Russian Federation}\\
\noindent $^{h)}${CEA, Centre d'\'Etudes Nucl\'eaires de Saclay, IRAMIS, 91191 Gif-sur-Yvette Cedex, France}\\
\noindent $^{i)}${Laboratoire Souterrain de Modane, CEA-CNRS, 1125 route de Bardonn\`eche, 73500 Modane, France}
\end{center}

\vspace{4pt}
\begin{abstract}
\small
Muon-induced neutrons constitute a prominent background component in a number of low count rate experiments, namely direct searches for Dark Matter. In this work we describe a neutron detector to measure this background in an underground laboratory, the Laboratoire Souterrain de Modane. The system is based on 1~m$^3$ of Gd-loaded scintillator and it is linked with the muon veto of the EDELWEISS-II experiment for coincident muon detection. The system was installed in autumn 2008 and passed since then a number of commissioning tests proving its full functionality. The data-taking is continuously ongoing and a count rate of the order of 1 muon-induced neutron per day has been achieved.
\end{abstract}

\begin{flushleft}
\end{flushleft}

\twocolumn
\section{Introduction}
\label{sec:intro}
A common challenge in rare-event searches is an accurate and precise understanding of all background components. A typical approach is based on a combination of Monte-Carlo simulations and dedicated measurements. In this work we present a detection system to study the production of muon-induced neutrons in an underground laboratory. The main goal is to perform a quantitative analysis of this background component in the specific environment of Laboratoire Souterrain de Modane (LSM) within the context of EDELWEISS-II~\cite{edw10,edw09dm} and EURECA \cite{kraus07} direct dark matter (DM) search experiments. Such experiments probe nature for weakly interacting massive particles (WIMP), as the most promising DM candidate, by means of nuclear recoils originating from a WIMP-nucleon elastic scattering. The expected interaction rate of WIMPs is extremely small, i.e. below 0.01 event per day and per kilogram of the target material, thus rising up the importance of the background knowledge. The recent developments in discrimination techniques of the DM detectors \cite{edw09id} allow a very powerful rejection of the electromagnetic background while neutrons reaching a detector with energies of the order of 1~MeV, in the case of EDELWEISS-II, can cause a signal potentially indistinguishable from the WIMP-nucleon scattering. The neutrons arise in an underground laboratory either from natural radioactivity having initial energies below 10~MeV, which can be effectively shielded, or being produced in a muon-induced cascade with energies up to several hundred MeV and able to travel far from a production vertex. In this paper we concentrate on studying the latter ones. The effect of the muon-induced neutron background can be reduced considerably by tagging the original muons with a muon veto around an experiment. However, the neutrons produced in the rock by muons passing outside of the setup and hence missed by the veto, can still reach the DM detectors. These neutrons can also lead to an \textit{in-situ} production of secondary neutrons via spallation, especially in a lead (Pb) material commonly used around the DM detectors for an efficient gamma attenuation. In addition, a fraction of muons can be missed due to a non-ideal muon veto efficiency and result in a neutron production inside the shielding, again predominantly in the Pb material.
 
In general, to measure the muon-induced neutron background is a challenging task due to its low intensity and its dependence on the actual experimental set-up. For instance, one could look for coincidences between the muon veto and the DM detectors but as the current data-taking of EDELWEISS-II shows, only few muon-induced events are identified in an exposure of several hundred kg.days~\cite{edw09dm, chantelauze09}. To perform corresponding Monte-Carlo simulations is not a simple issue either, since many parameters have to be taken into account, e.g. the muon energy spectrum and the flux in an underground laboratory, production of neutrons by hadronic and electromagnetic muon-induced showers, material dependency of this production, three dimensional topology of interactions. As reviewed in \cite{araujo05, mei06}, measurements of neutron production rates in liquid scintillator at different laboratory depths can be well described by most frequently used packages, GEANT4~\cite{geant4} and FLUKA~\cite{fluka}, which are in agreement with each other within $\leq$30\% at the muon energies specific for an underground laboratory. However, the calculated neutron yield for Pb and an ordinary rock can vary up to a factor of 2 between the packages, being lower in GEANT4 \cite{araujo05}. Experimental neutron production rates in Pb measured in \cite{gorshkov74, bergamasco73} and scaled to the same mean muon energy appear to be inconsistent within a factor of 3 among each other and may suggest that both simulation packages underestimate the neutron yield in heavy targets \cite{araujo05, wulandari04}. On the other hand, recent studies performed in \cite{araujo08} with an existing experimental setup though not fully optimized to the specific needs of muon-induced neutron detection result in a measured neutron yield in Pb being 1.8 (3.6) times smaller than predicted in GEANT4 (FLUKA). Overall, these measurements give a spread of neutron production rate in Pb of nearly one order of magnitude which can affect the sensitivity of DM experiments by the same factor.

In order to measure the muon-induced neutron background and to develop the corresponding background model optimized for the LSM laboratory hosting the EDELWEISS-II experiment and the future one, EURECA, we have designed and installed a \textit{dedicated} neutron counter. Having the neutron detector \textit{in-location} of present and planned DM set-ups allows to avoid several uncertainties in the simulations, e.g. arising from calculations of the muon energy spectrum and the muon flux in the laboratory. The goal is to investigate the muon-induced neutron yield in Pb either with or without tagging the incident muon to an accuracy of better than 25\% with equal statistical and systematic uncertanties expected. One should also note here, that the EDELWEISS-II strategy on neutron background investigation is complemented by continuous monitoring of the thermal neutron flux in the vicinity of the DM experiment by means of $^3$He detectors \cite{rozov10}. Another new experiment to study muon-induced neutron flux deep underground was proposed by other authors in \cite{hennings07}. The investigation of the neutron background is, in fact, not limited to the field of dark matter and is relevant, for instance, for experiments looking for neutrinoless double beta decay, e.g. NEMO-3 \cite{nemo305} and SuperNemo \cite{supernemo09} with their location in LSM too. Furthermore, the results of our studies of muon-induced neutrons are as well applicable to other underground laboratories.

The structure of the paper is as follows: in Sect.~\ref{sec:muons}, a general overview of muon interactions and production of secondary neutrons is given, emphasizing the role for DM search. Sect.~\ref{sec:setup} describes the details of the neutron detector installed in LSM, including the data acquisition system, LED-based control of the liquid scintillator and a safety monitoring. Results of commissioning runs demonstrating the performance of the system are discussed in Sect.~\ref{sec:commiss}. Sect.~\ref{sec:meas} presents preliminary data from first physics runs as proof-of-principle. In Sect.~\ref{sec:sum}, we conclude with a summary and an outlook of the anticipated data taking necessary to get full physics results.

\section{Muon induced neutrons}
\label{sec:muons}
\begin{figure}
\centering
\resizebox{0.45\textwidth}{!}{%
  \includegraphics{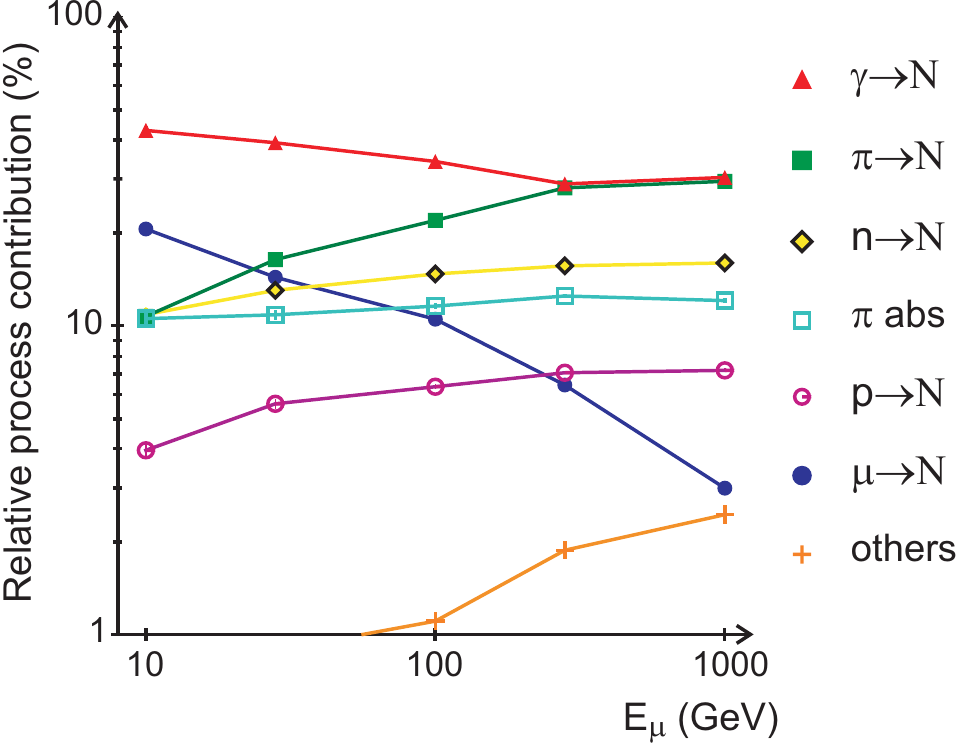}
}
\caption{Relative contribution of individual processes to the total neutron production yield as a function of the muon energy in C$_n$H$_{2n}$, GEANT4 simulations. Figure is adopted from \cite{horn07}, the lines are drawn to guide the eye.}
\label{fig:mu-n-production}  
\end{figure}

Atmospheric muons are mainly produced in decays of charged pions and kaons created in an interaction of primary cosmic rays with the atoms of the Earth's atmosphere. The energy spectrum of muons extends above 1~TeV while the mean energy is about 4~GeV at the ground~\cite{pdg08}. The overall angular distribution of muons at the sea level is $\propto$~cos$^2\theta$ ($\theta$ being the zenith angle), at lower energy it becomes increasingly steep, while at higher energy it flattens, approaching a 1/cos$\,\theta$ distribution for $E_{\mu}\gg115$~GeV and $\theta<70^{\circ}$~\cite{pdg08}. The total muon energy spectrum in the underground lab can be calculated rather well in the assumption of flat overburden \cite{mei06,horn07} and it becomes harder, e.g. $\left\langle E^{LSM}_\mu\right\rangle=260-290$~GeV. To get a corresponding angular distribution for an underground site, a mountain profile has to be taken into account and thus it becomes laboratory specific. When passing through the rock, the muons lose their energy in four main processes: ionization (including the production of high-energy $\delta$-electrons), production of e$^-$e$^+$-pairs, bremsstrahlung and deep inelastic scattering (DIS). These processes naturally also lead to a development of electromagnetic and hadronic showers. Therefore, under the muon-induced background one has to consider, in general, all particles created both directly by original muons and in secondary cascades. As it was already explained in the introduction, neutrons constitute the most dangerous background component for direct DM searches among other muon-induced particles. GEANT4 simulations show (Fig.~\ref{fig:mu-n-production}) that most of the neutrons are created in the secondary cascades, predominantly by photonuclear interaction, pion and neutron inelastic scattering, pion absorption at rest and proton inelastic scattering~\cite{araujo05,horn07}. The contribution of direct muon nuclear interaction being the second most important one at low muon energies decreases with higher energies because of larger electromagnetic and hadronic cascades. Neutrons produced in electromagnetic interactions have typically low energy and can be effectively absorbed. In contrast, the hadronic showers and especially high energy muon DIS reactions lead to a production of energetic neutrons which can travel far away from the creation point, reach the DM set-up, penetrate the entire shielding and cause nuclear recoils in the DM detectors either directly or by a neutron spallation in the high-Z material surrounding the experimental setup. It also has to be noted that neutron production generally increases with the muon energy and the average atomic weight of the material, e.g. in Pb there will be $\sim$100 times more neutrons created compared to C$_n$H$_{2n}$ and $\sim$10 times more compared to rock or concrete of the same thickness \cite{kudryavtsev03,araujo05,mei06,horn07}. To model the effect of the muon-induced neutrons in application to DM searches, it is thus essential to take into account all the processes aforementioned, track each single particle through different materials down to an energy deposit in DM detectors on an event-by-event base. Monte-Carlo simulations based on the GEANT4 toolkit (version 8.1.p01) were developed in \cite{horn07} and included all these aspects, profile of the mountain, full geometry of the EDELWEISS-II experiment and topology of an event. The EDELWEISS-II set-up, described elsewhere \cite{edw10}, hosts in a single cryostat an array of Ge bolometers used as DM detectors. The cryostat is surrounded by 20~cm lead shield, followed by 50~cm polyethylene and a modular muon veto \cite{chantelauze09, chantelauze07}. As the simulations show \cite{horn07}, most of neutrons leading to energy deposit in Ge bolometers are due to secondary production in the Pb shield near the cryostat. The rate of this background for the optimized EDELWEISS-II set-up can be as low as $\Gamma\sim10^{-5}$/kg/day once the anti-coincidence with the muon veto and with other bolometers is required but crucially depends on the shielding concept and the muon veto efficiency. As follows from these simulations and the ones described in \cite{mei06}, these muon-induced neutrons are not a matter of concern for the current phase of EDELWEISS-II aiming for the sensitivity of 10$^{-8}$~pb for the spin-independent WIMP-nucleon scattering cross-section but may become so for the next phase and especially for the next generation DM experiments with the goal of 10$^{-10}$~pb, such as EURECA. With the neutron counter described here we intend to gather statistically significant data for validation and further development of our background model \cite{horn07}.

\section{Experimental set-up}
\label{sec:setup}
\subsection{Measurement principle}
\label{sec:setup.intro}
The LSM underground lab, where the designed neutron detector is installed, is situated in the Fr\'ejus road tunnel in the French-Italian Alps and has an overburden shielding of 4850~m.w.e. This reduces the muon flux down to about 5~/m$^2$/day~\cite{berger89}, i.e. by more than a factor of $10^6$ comparing to sea level. Such a low rate naturally leads to a low flux of muon-induced neutrons and requires a relatively large size of the instrument. To measure the neutron yield originated from well-identified muons interacting in Pb, the developed neutron counter (NC) has a layer of Pb and its read-out system is coupled to the muon veto system of the EDELWEISS-II experiment for a good tagging of incident muons. In order to achieve an efficient detection of neutrons, a Gadolinium-loaded liquid scintillator is used. The primary trigger in this case is associated with the muon itself and following triggers correspond to secondary neutrons predominantly created in Pb and captured thereafter on Gadolinium (Gd). A measurement of the neutron production rate without demanding a primary muon tag can only be done by requiring a multiple trigger occurrence in the NC within a predefined acquisition window ($t_{DAQ}\approx60\;\mu$s). Together with an appropriate energy threshold this allows to effectively reject accidental background. For these events the primary trigger arises from any particle produced in muon cascades. For instance, high energy gamma's can first lead to Compton scattering in the NC scintillator volume and only then produce secondary neutrons in the Pb layer via photonuclear reaction. High energy neutrons could create a prompt recoil proton at first and then generate secondary neutrons being characterized as a multiple neutron capture event. In this mode we study the overall neutron production rate without identifying the primary particle. The events with multiplicity 2, i.e. a primary trigger and one secondary, may in part contain a contribution of correlated backgrounds, e.g. the fast neutrons originating from natural radioactivity or $\beta$-decays of $^{214}$Bi followed by an $\alpha$-decay of $^{214}$Po: The flux of the ambient neutrons is $10^{-6}$~neutrons/cm$^2$/s above 1~MeV \cite{lemrani06}, the energy spectrum extends to 10~MeV but is strongly suppressed above 4~MeV \cite{lemrani06, chazal98}. Nevertheless, these neutrons can still lead to recoil protons above the NC threshold and then be captured in the scintillator. $^{214}$Bi comes from the Uranium-chain, undergoes the $\beta$-decay with the endpoint energy of 3.27~MeV to $^{214}$Po which then $\alpha$-decays with a half-life of 164.3~$\mu$s and E$_\alpha = 6.6, 6.9, 7.68$~MeV. Thus both $\beta$- and $\alpha$-particles can deposit energy above the NC threshold within the acquisition window of $t_{DAQ}\approx60\;\mu$s. This implies that events with one secondary hit need special attention in the analysis while the effect of both background components is negligible for higher multiplicities in our studies.

\subsection{General overview}
\label{sec:setup.all}
\begin{figure}
\centering
\resizebox{0.45\textwidth}{!}{%
  \includegraphics{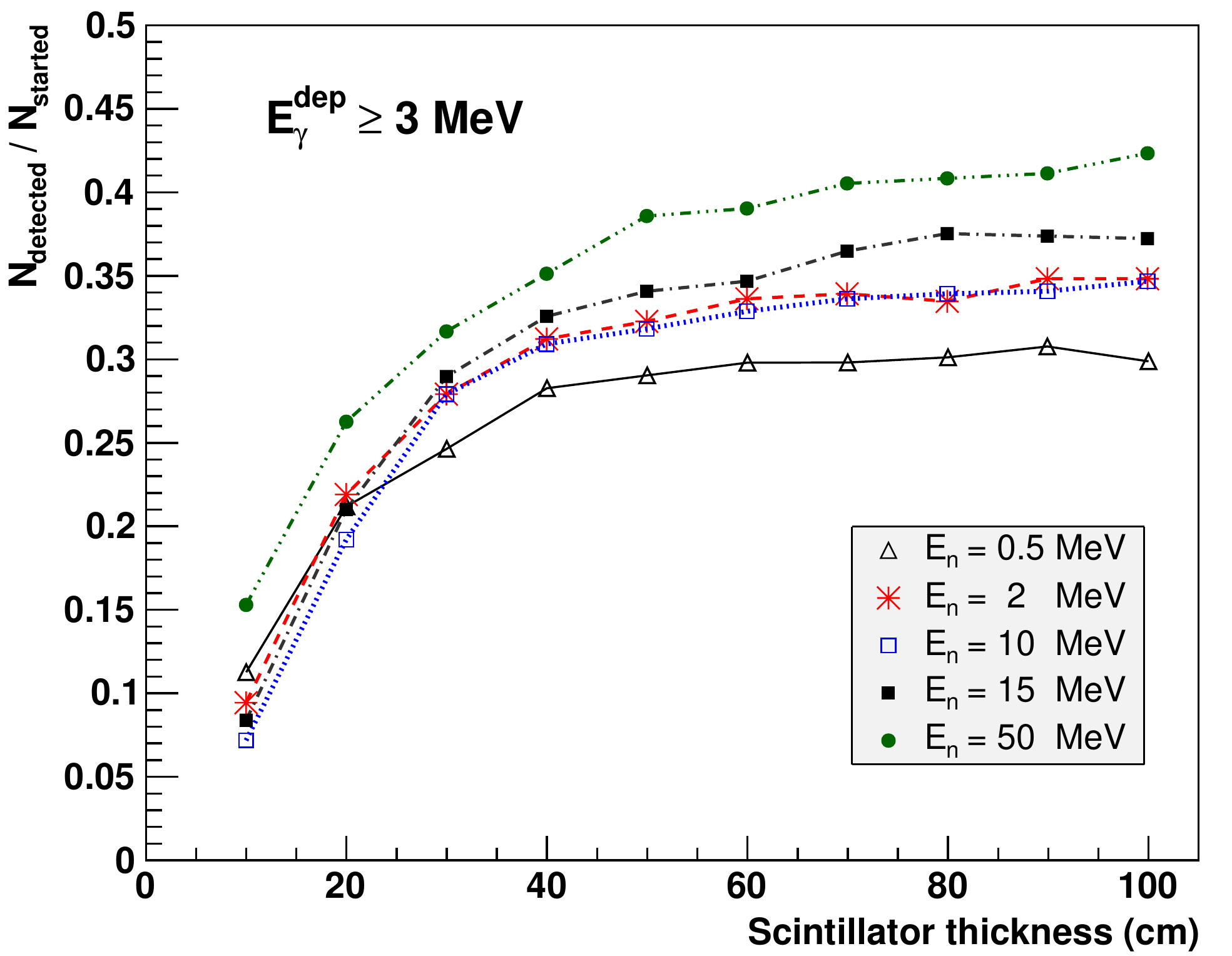}
}
\caption{Neutron detection efficiency, i.e. number of registered neutrons vs. number of incident neutrons, as a function of scintillator thickness (GEANT4 simulation). Neutron is considered to be registered when it is captured and the following $\gamma$'s deposit in the scintillator a sum energy above 3~MeV in this case. Neutrons start isotropically with different energies (indicated on the plot) and equally distributed over a horizontal bottom surface of the detector.}
\label{fig:g4-scint}  
\end{figure}

\begin{figure*}
    \begin{minipage}[c]{0.73\linewidth}
        \centering
        \resizebox{0.95\textwidth}{!}{%
           \includegraphics{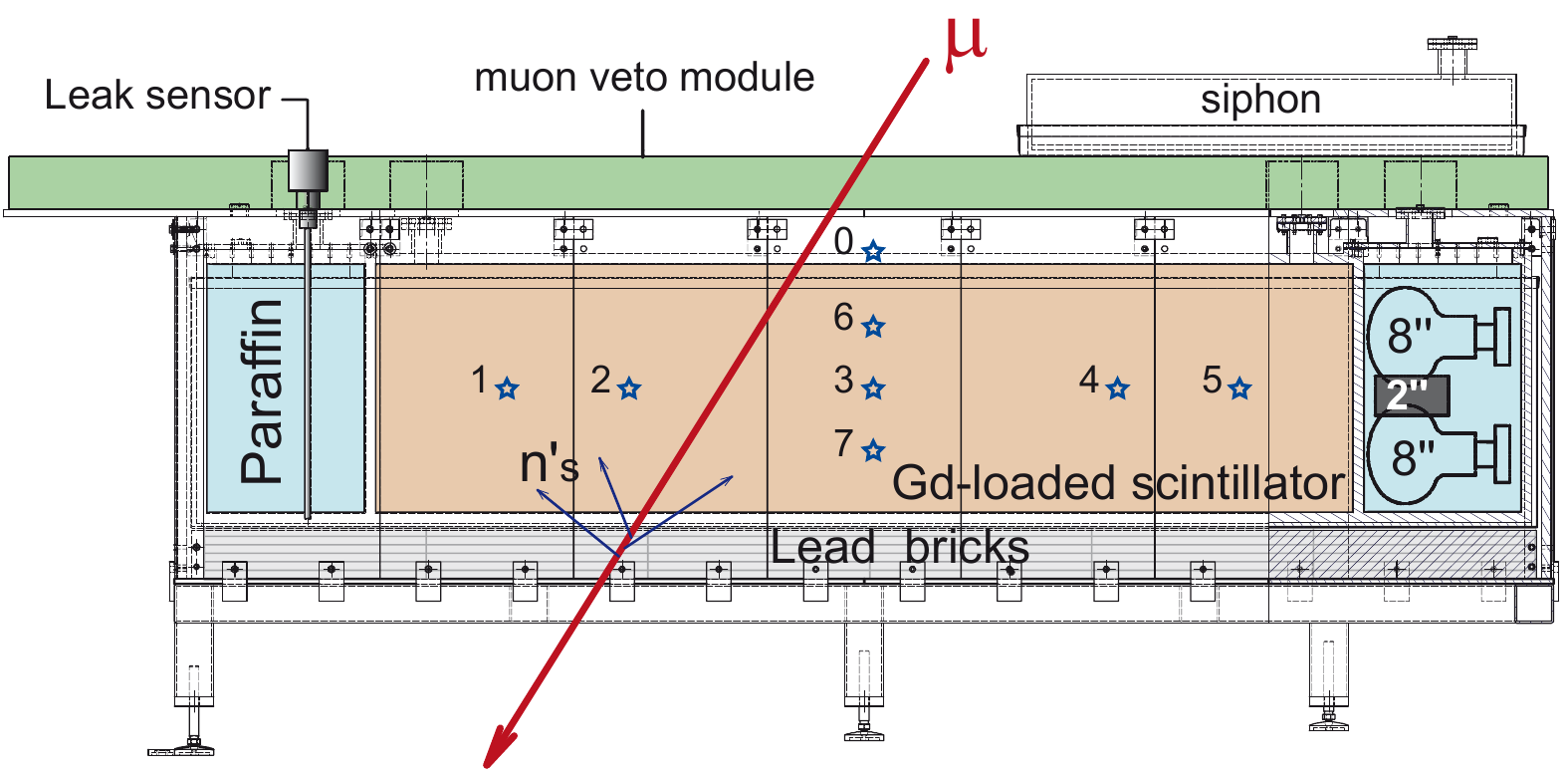}
        }
    \end{minipage}
    \begin{minipage}[c]{0.25\linewidth}
        \centering
        \resizebox{0.95\textwidth}{!}{%
           \includegraphics{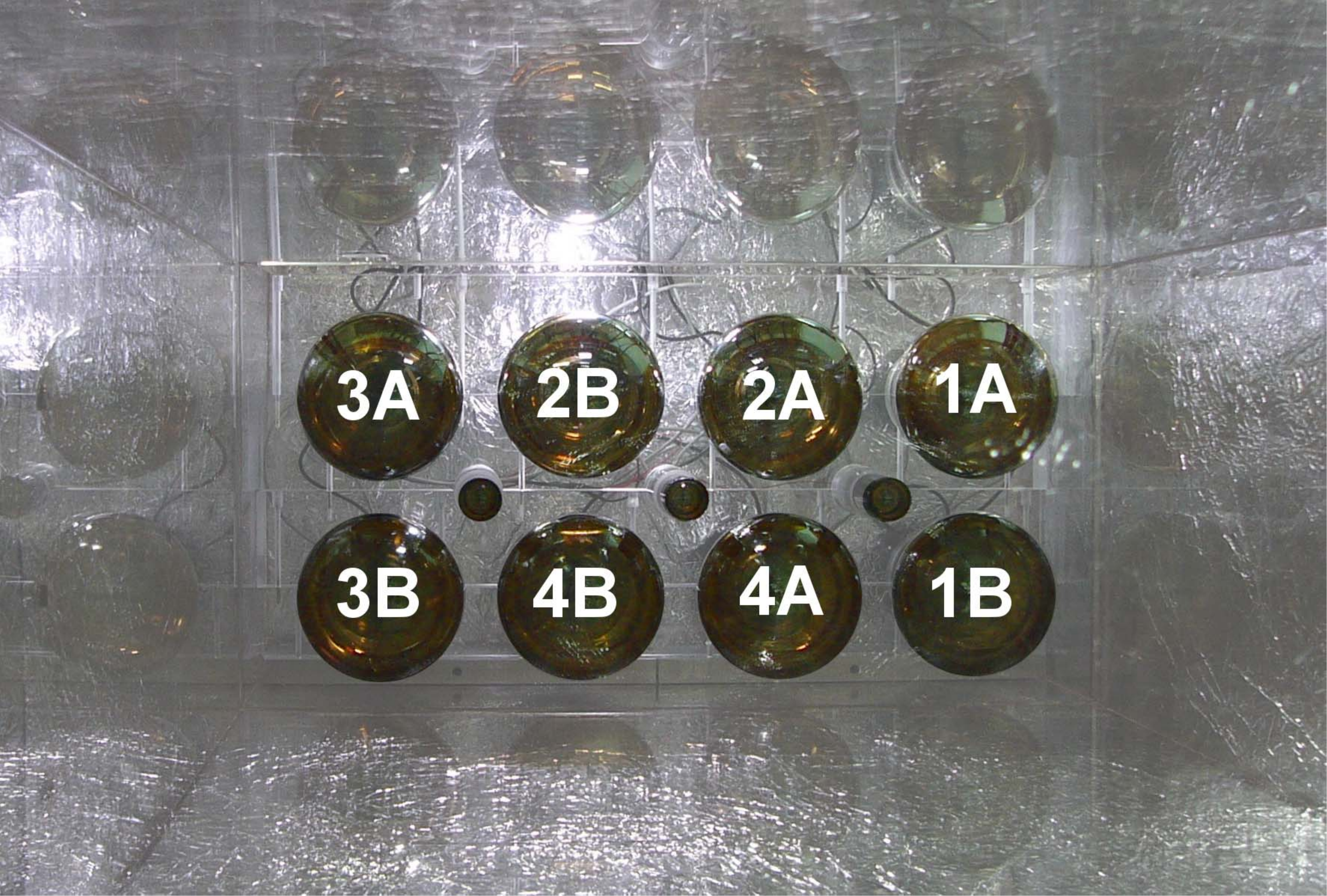}
        }
    \end{minipage}        
    \caption{\textit{Left}: general layout of the neutron counter detector, positions of LED's are indicated by stars. \textit{Right}: a view from inside of the installed but not yet filled scintillator chamber, 8 of the 8-inch and 3 of the 2-inch PMTs are seen. Numbers indicate different PMT groups consisting of two 8-inch PMTs (A,B) each.}
    \label{fig:setup}
\end{figure*}

The core of the detector consists of 1~m$^3$ of liquid scintillator (50x100x200~cm$^3$) loaded with 0.2\% of Gd by weight, St.~Gobain Bicron BC525. The neutron capture on Gd has the largest cross-section ($\sigma_\gamma=2.54\cdot10^5$~b for $^{157}$Gd, 15.65\% of natural abundance \cite{toi98}) compared to other stable elements and results in several $\gamma$'s with a summed energy of 8~MeV. Adding 0.2\% of Gd in organic scintillator leads to a capture time, $\tau_{capture}$, of about $17\;\mu$s while a typical time needed for neutrons to thermalize is defined by the proton density in the liquid and is $\sim$10~$\mu$s (see Sect.~\ref{sec:commiss.calib}). Underneath the detector, a 10-cm thick layer of lead bricks is put for studies of the neutron yield. The thickness of this layer was chosen by considering the process of secondary neutron production, the probability for these neutrons to reach the scintillator and the mechanical stability of the instrument. The geometry of the NC scintillator volume was optimized for the efficiency of neutron detection with specially adapted Monte-Carlo simulations based on GEANT4 (version 8.1.p01) and taking into account space and security restrictions of the underground laboratory. In the simulations it was found that most of the secondary neutrons are captured  after $\sim$20~cm of penetration into the scintillator. The registration efficiency of gammas produced by Gd(n,$\gamma$) saturates once a 50-cm thick system is realized (Fig.~\ref{fig:g4-scint}), thus defining the thickness of the instrument. However, this efficiency is significantly less than 100\% because of a partial escape of neutron-capture gammas as a consequence of a limited neutron penetration into the scintillator, as well as due to an applied energy threshold. As an example, for $E_n=10$~MeV and a threshold of 3~MeV it reaches a value of $\sim$33\%. For higher neutron energies, the inelastic scattering process (n,$\;$2n) starts to play a role, thus increasing the detection efficiency. Due to the partial $\gamma$ escape it is obvious that, instead of the $\gamma$-peak at 8~MeV sum energy originating from the neutron capture on Gd, a continuous spectrum extending to 8~MeV is expected. The detector volume is divided in three compartments: a central one containg the 1~m$^3$ of scintillator, and two smaller ones on either side to hold the photomultipliers and filled with highly transparent mineral oil (paraffin) (Fig.~\ref{fig:setup}). There are in total 16 PMTs (8 on each side) of Hamamatsu R5912 type$\,$\footnote{these PMT's are called \textit{neutron PMT's} in the following}, 8-inch diameter, the gain and the read-out electronics of which are optimized for the neutron capture events. In addition, 6 smaller PMTs (3 on each side) of Philips VALVO XP 2262 type$\,$\footnote{these PMT's are named correspondently \textit{muon PMT's}}, 2-inch diameter, are installed in between of these 8-inch photomultpliers. Since a typical muon signal is well above the dynamic range of the PMT's used for the neutron registration, the 2-inch PMT's are operated at a gain adapted to the typical muon energy deposit of the order of 100~MeV. The entire container holding all liquids is made of a transparent plexiglass. A collection of scintillation light thus benefits from the total reflection at the outside surfaces. In order to avoid a pressure build-up possible due to a temperature variation, an expansion bag filled with an argon gas is connected to the scintillator volume through a siphon system. The plexiglass chamber is wrapped with an aluminum foil to further improve the light collection. It is then placed in an aluminum vessel as a safety container. Finally, the system is surrounded by iron plates (excluding top plane) to reflect a fraction of neutrons back to the scintillator. To allow a coincidence measurement of the incident muon, the NC is coupled electronically with the existing muon veto of the EDELWEISS-II experiment. This veto system ($\mu$Veto) is constructed from 42 plastic scintillator modules (BC-412 type from St.Gobain) with a total of 100~m$^2$ of surface. It is further extended by an additional module (5x65x315~cm$^3$) placed on top of the neutron detector and the 2-inch PMTs of the NC, described above.  The complete set-up is positioned right near the western wall of the EDELWEISS muon veto system.  

\subsection{Data acquisition}
\label{sec:setup.daq}
\begin{figure*}
\centering
\resizebox{0.8\textwidth}{!}{%
  \includegraphics{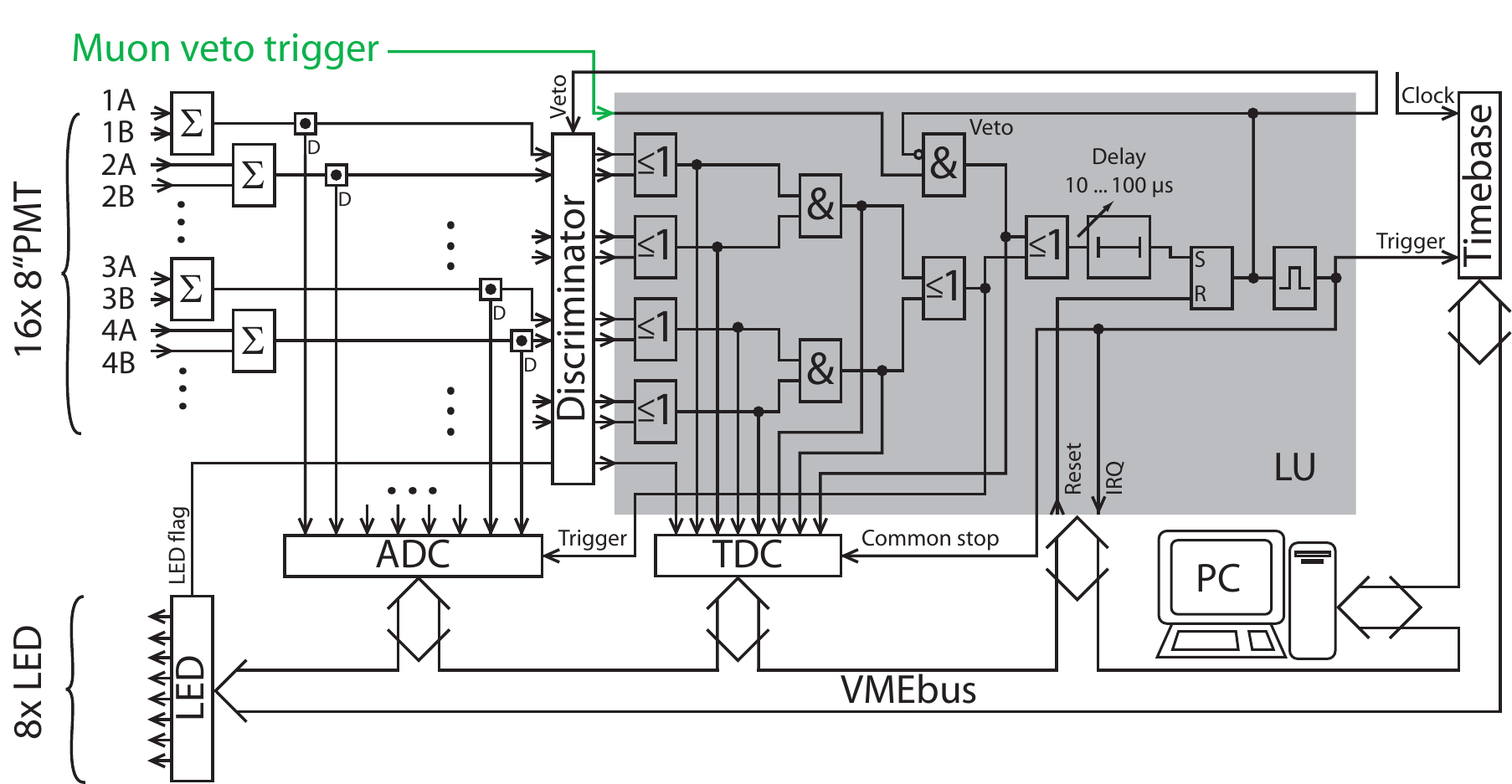}
}
\caption{General layout of the VME-based data acquisition, see Sect.~\ref{sec:setup.daq} for details.}
\label{fig:daq}  
\end{figure*}

As was described in Sect.~\ref{sec:setup.intro}, two event signatures for muon-induced neutrons are considered: \textit{i}) A high energy muon triggers either the muon system only or the muon veto and the neutron counter together. A delayed signal due to a neutron capture is then looked for in the NC; \textit{ii}) Multiple trigger occurrence happens in the NC detector within a defined time window ($t_{DAQ}\approx60\;\mu$s). The developed data acquisition system (DAQ) accounts for both of these possibilities and allows to configure two different measurement subsystems separately optimized for cases \textit{i}) and \textit{ii}). It is based on VME electronics, integrated into the $\mu$Veto DAQ and uses specially built software in a Linux environment. First, in order to enhance the light collection two neighboring \textit{neutron PMT's} (denoted A and B, Fig~\ref{fig:setup}(\textit{right}) and Fig.~\ref{fig:daq}) are combined into one group with individual high voltage supply. Thus, already summed signals are used for an event selection and one measurement subsystem consists of 4 PMT groups, 2 on each side. The signals of each group are split and sent to a discriminator, CAEN V895, as well as with a delay to a flash analog-to-digital converter (ADC), CAEN VX1720, 250~MS/s. Once a threshold is exceeded, logic ECL signals generated by the discriminator are delivered to a in-home developed \textit{Logic Unit} (LU), based on field-programmable gate array (FPGA). This LU takes care of the final acceptance of the event, where the event is considered to happen when at least two opposite PMT groups related to the same measurement subsystem get a light signal above threshold. In the LU module this is realized as follows. Signals from two neighboring PMT groups primarily pass through an \textit{OR}-logic. Two \textit{OR}-units having as an input the signals of opposite PMT groups are then wired to an \textit{AND}-logic, which are thus 2 in total and define our two measurement subsystems. A coincidence on any of these two \textit{AND}-units gives a trigger to the flash ADC to convert an analog signal. After a defined delay of $t_{DAQ}\approx60\;\mu$s, the LU generates a common stop signal, i.e. ends the acquisition cycle, actuates the DAQ software and withdraws a time stamp of the triggering event from an external clock (10~$\mu$s precision). The latter allows to get a direct correspondence between NC and $\mu$Veto events. Each of the described above \textit{OR}- and \textit{AND}-outputs is connected to a time-to-digital converter (TDC), LeCroy 1176, which allows to have a time information of individual hits within the event with the resolution of 1~ns. The flash ADC allows to reconstruct an energy signal of every individual hit within $t_{DAQ}$ without creating a dead time. In addition, the LU card is triggered also in case of a muon-like event registered by the $\mu$Veto. While the DAQ software records an acquired event, the VME bus is locked to avoid distortion of the data. Once the event is processed, the VME bus is reset and the system is ready for a next event. The dead time to treat an event was measured with a pulse generator and was found to be 6.76~$\pm$~0.08~ms, while the typical raw count rate of the NC is 10~Hz.

\subsection{Monitoring systems}
\label{sec:setup.monitor}
When dealing with liquid scintillators, especially Gd-loaded ones, it is important to control their aging as it can influence the registration efficiency with time. For this purpose the NC is equipped with a dedicated system of light-emission-diodes (LED) of 425 nm wavelength, corresponding to the maximum emission of the scintillator. There are 8 LED's in total: seven located on the same flank of the NC and one, marked "0", in the center of the top plain (Fig.~\ref{fig:setup}(\textit{left})). Their control is performed from the DAQ computer via a specifically designed VME-based module. It is possible to adjust the voltage of each LED and an individual time of firing, while the duration of a light pulse is fixed by electronic components resulting in a PMT signal of about 10~ns (FWHM). The corresponding electronics is based on the LED calibration system designed for the Pierre Auger Observatory \cite{auger04}. The control module also generates at the time of LED ignition a logic pulse, or \textit{LED flag}, which is then given as an input to the TDC of the DAQ (Fig.~\ref{fig:daq}). This flag allows to clean out the data from the LED entries while searching for the muon-induced events in an offline analysis. The LED's are typically ignited three times per day with a frequency of 1~Hz in a predefined cycle.

Another issue of the liquid scintillator used is its pseudocumene base, which requires a careful handling. A safety system was thus designed and installed. It contains several vapour detectors, two liquid leak sensors, temperature measurement outside and inside of the system as well as monitoring of the siphon level. All parameters are accessible online through a web-based interface. More details on this slow control monitoring can be found in \cite{kozlov08}.

\section{Commissioning}
\label{sec:commiss}
The NC detector was installed in LSM in autumn 2008. After a commissioning period, data-taking started in early 2009. 
\subsection{Control of the scintillator transparency}
\label{sec:commiss.scint}
\begin{figure}
\resizebox{0.47\textwidth}{!}{%
  \includegraphics{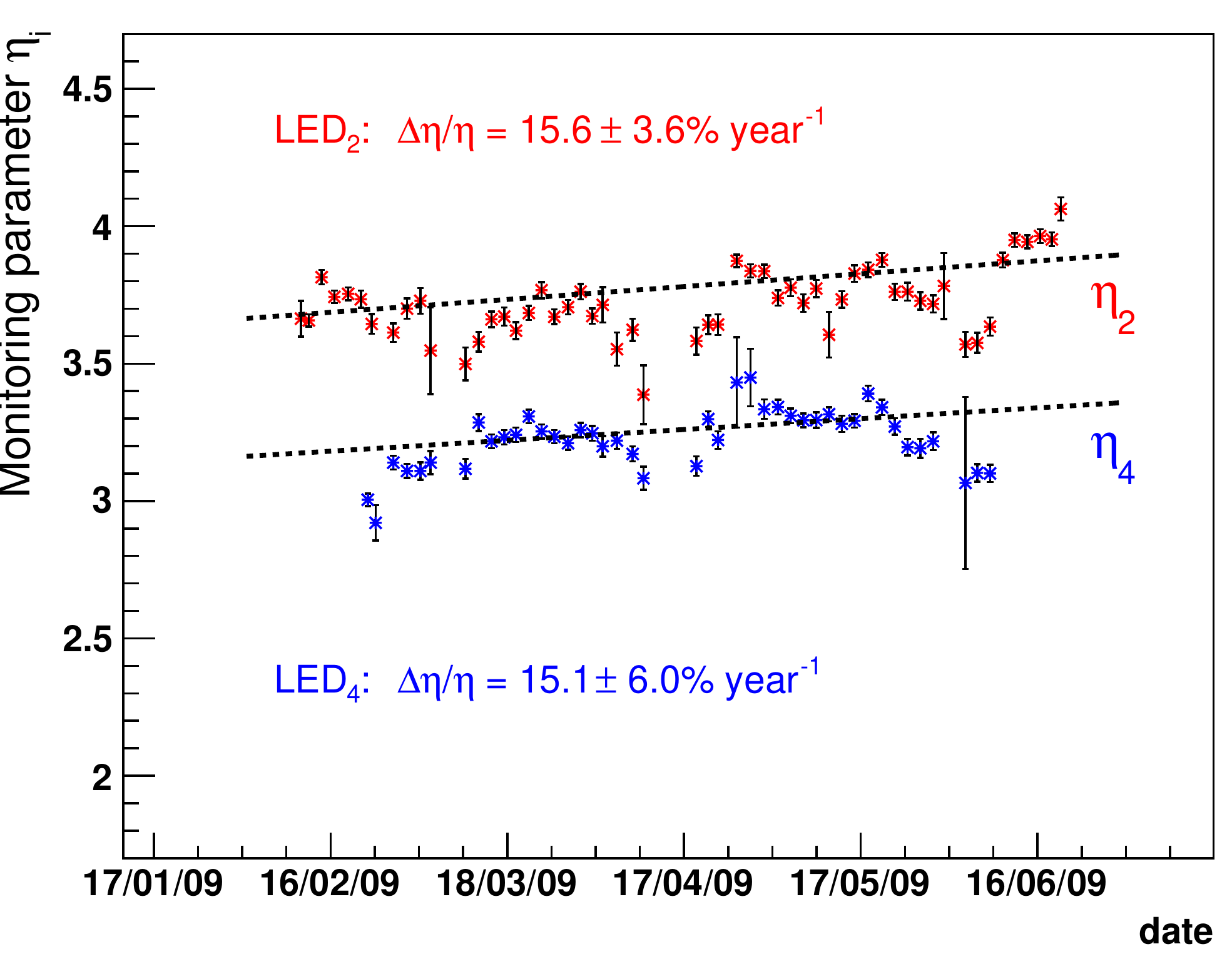}
}
\caption{Long term behaviour of the scintillator transparency: 4-months trends of the $\eta$ ratio for two non-central LED's: LED$_2$(\textit{top}) and LED$_4$(\textit{bottom}) (see Sect.~\ref{sec:commiss.scint} for details). Both data sets are fit independently with a linear function. }
\label{fig:scintLED}  
\end{figure}

Data acquired each day with the LED's are analyzed to monitor the light properties of the scintillator in the following way: one of the central LED's, e.g. LED$_0$, is used first to normalize the gain of the PMT's and to account for any temporal fluctuations of PMT's and electronics. Then for the measurement performed with a non-central LED$_i$, a ratio of ADC signals of ``near'' PMT's, i.e. those located closer to the chosen LED$_i\;$, and ``far'' PMT's is calculated:
\[
	\eta_i = \frac{ADC^{(i)}_{near\;PMT's}\:/\:ADC^{(0)}_{near\;PMT's}}{ADC^{(i)}_{far\;PMT's}\:/\:ADC^{(0)}_{far\;PMT's}}\:,\quad i=1\:..\:7
\]
Any changes of $\eta_i$ over time can thus be attributed only to changes in light transport, predominately in the transparency of the scintillator. The parameters $\eta_{i}$ are monitored on an everyday basis and 4-months trends measured with two non-central but symmetrically-positioned LED's, LED$_2$ and LED$_4$ (Fig.~\ref{fig:setup}(\textit{left})), show consistent relative changes of the $\eta$-ratio (Fig.~\ref{fig:scintLED}). Analysis of the combined data yields a degradation of 15.5$\pm$3.1$\,$\% per year as the weighted average. A degrading light signal can be compensated by increasing the PMT sensitivity accordingly, e.g. by raising the high voltage of the PMTs. In such case this degradation is acceptable for an expected run period of the NC of 2-3 years. The fluctuations of the $\eta$-ratio seen in Fig.~\ref{fig:scintLED} are typically correlated with variations of the scintillator temperature.
The same $\eta$-ratio is also reconstructed in GEANT4 simulations of the NC with light propagation included. The aluminum foil is described with the \textit{groundfrontpainted} model for a diffuse reflection. A value of 0.9 for Al reflectivity is taken, optical properties of the scintillator and the plexiglass are considered. A difference between the absolute $\eta$-values of two measurements as seen in Fig.~\ref{fig:scintLED} can be reproduced in the simulations and can be explained by a misalignement of the LED's and the directional property of the LED light. At present, this simulation agrees with the measurement within 20\% and further improvement of the light tracking model is ongoing.

\subsection{Identification of muon events}
\label{sec:commiss.mu}
\begin{figure}
\resizebox{0.47\textwidth}{!}{%
  \includegraphics{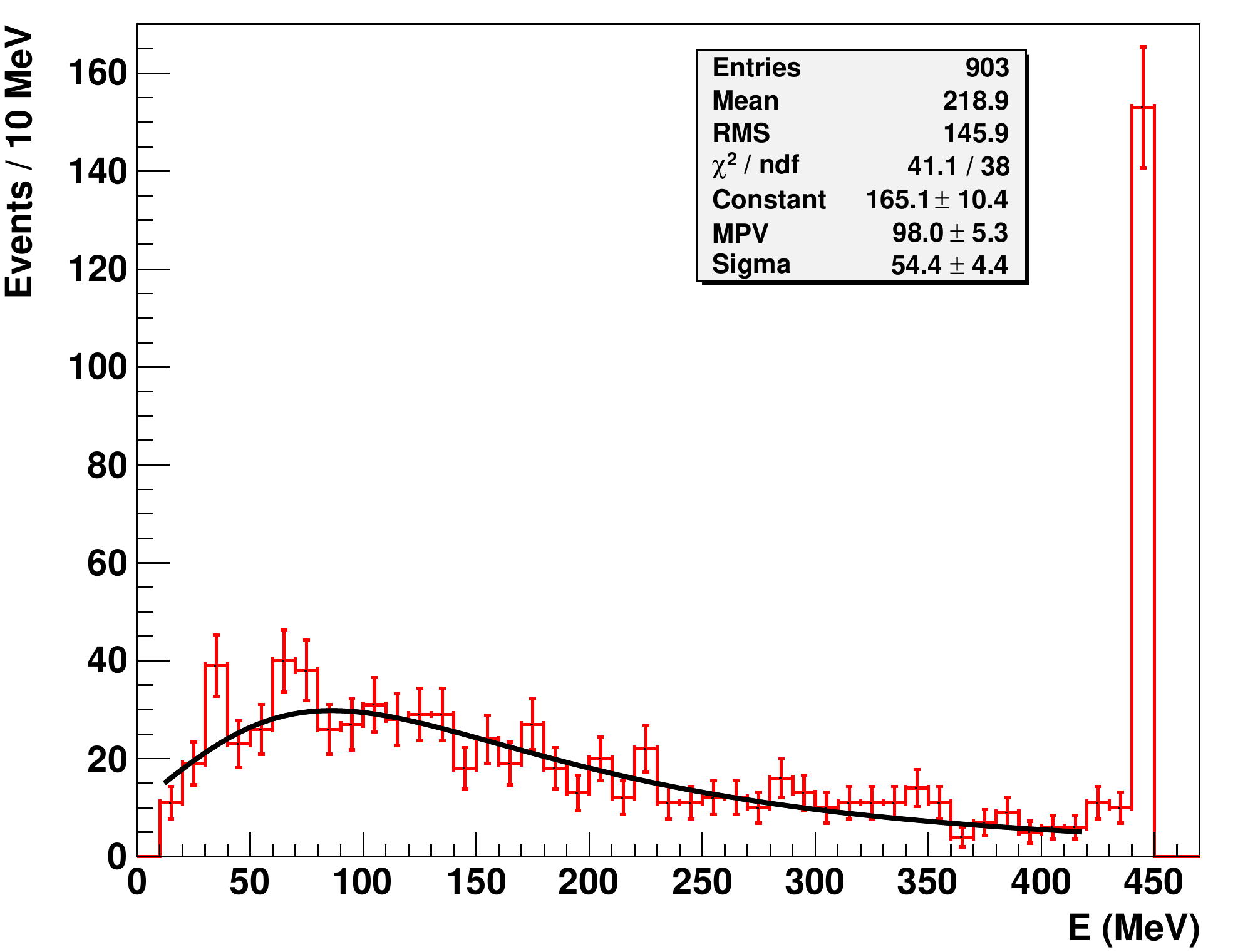}
}
\caption{Energy deposition measured with the \textit{muon PMT's} of NC for the events in coincidence with the $\mu$Veto module on top of the NC, live time is 177.9 days. Last bin contains events with energy deposition of E$\geq$440~MeV (overflow bin). Landau distribution is fitted to the data and calibrated with MC simulations of muons passing through the NC.}
\label{fig:nc-muons}  
\end{figure}

Since the described detector is accompanied by two complementary systems to identify in particular a muon crossing the NC ($\mu$Veto plastic scintillator module on top and the \textit{muon PMT's} inside of the NC), their performance has to be checked as well. The average energy of muons passing the mountain overburden is $\left\langle E^{LSM}_\mu\right\rangle=260-290$~GeV, thus one expects energy losses in a 50~cm thick detector following a Landau distribution. By selecting events registered by both \textit{muon PMT's} and the top $\mu$Veto module and reconstructing their energy deposit in the volume of the neutron detector, one finds a good agreement with a Landau expectation (Fig.~\ref{fig:nc-muons}). This spectrum is rather broad due to a superposition of various track lengths ranging from more than 50~cm to very short ones for grazing muons in the NC. The count rate of events that trigger both systems is $\Gamma_\mu=5.1\pm0.2$~/day, from which one can deduce the muon flux in LSM using the geometrical area and the solid angle of the system. The geometrical overlap is S=1.3~m$^2$, the solid angle is less than 2$\pi$ due to a 15~cm gap between the $\mu$Veto module and the liquid scintillator level in the NC due to the mechanical support. In addition, a fraction of muons is missed due to short track lengths of grazing muons leading to energy deposits below the threshold (see the low energy cutoff in Fig.~\ref{fig:nc-muons}). All together, this results in an effective reduction of the measured muon flux estimated to be as large as 30\%. Correcting for this, we get an estimated muon flux of 5~/m$^2$/day which is in agreement with an earlier measurement of $\Phi_\mu =4.98\pm0.09$~/m$^2$/day \cite{berger89}.

\subsection{Verification of neutron detection}
\label{sec:commiss.calib}
\begin{figure}
\resizebox{0.47\textwidth}{!}{%
  \includegraphics{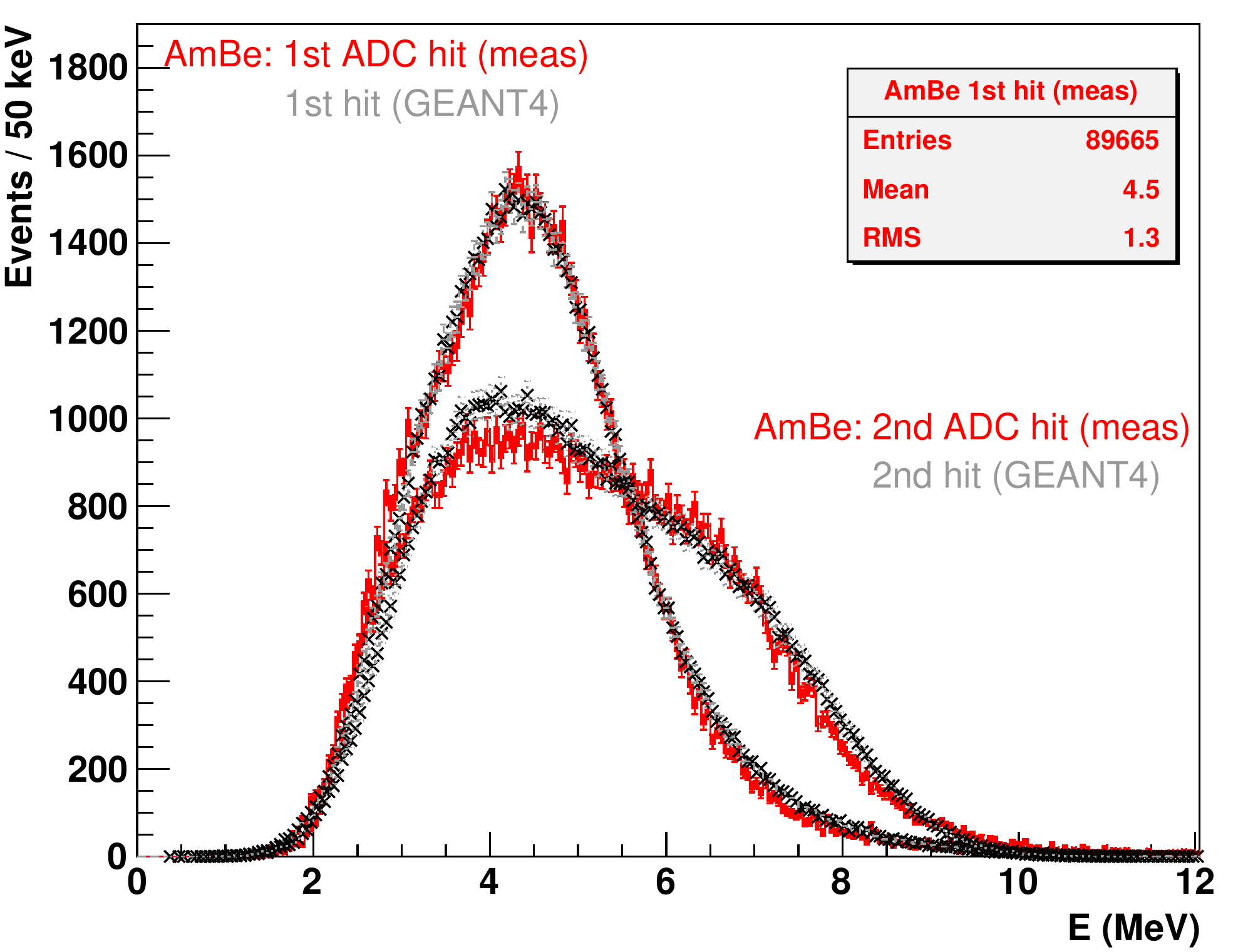}
}
\caption{Energy spectra acquired with AmBe source (spectra of the first and second ADC hits are shown) (\textit{solid lines}). Both ADC hits of AmBe measurement are compared with corresponding GEANT4 simulations (\textit{crosses, dashed line})(see text for details).}
\label{fig:calib}  
\end{figure}
\begin{figure}
\resizebox{0.47\textwidth}{!}{%
  \includegraphics{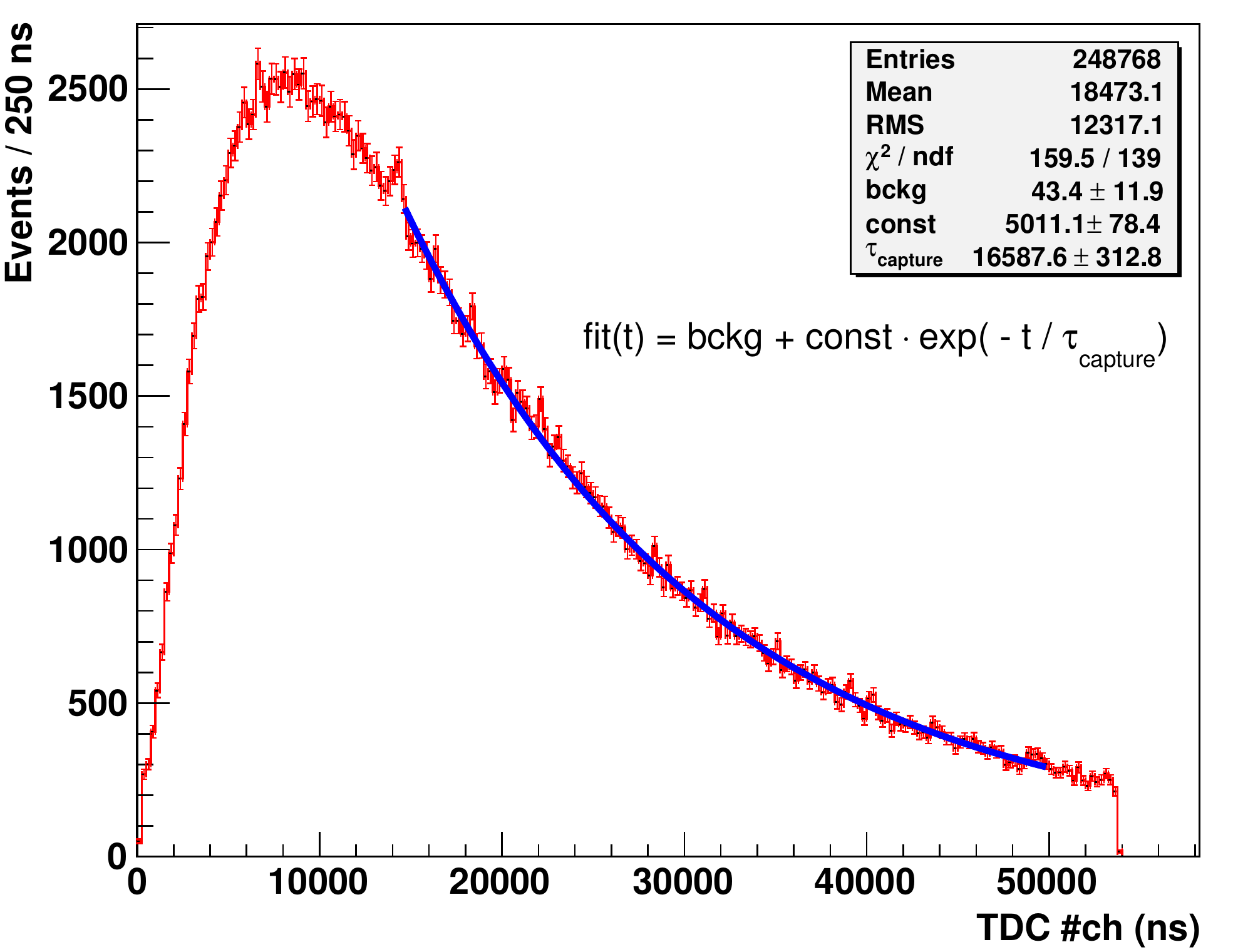}
}
\caption{AmBe measurement: time difference between a start of the event (first hit) and any following TDC hit within the acquisition time. The falling edge is fit with an exponential function and a flat background component. }
\label{fig:AmBe}  
\end{figure}

An energy threshold of the NC has to be set such that most of an intrinsic background is excluded but the registration efficiency for $\gamma$'s after the neutron capture is still high enough. Hence, this value was chosen to be around 2.5~MeV in the center of the detector. For the measurement of the neutron yield due to untagged muons the energy threshold is set to an even higher value of 3.5~MeV to further suppress accidental and correlated background. In order to put a proper value of the electronic threshold and to tune the gain of \textit{neutron PMT's}, a $^{60}$Co source was used first. Two gammas of 1173.2~keV and 1332.5~keV are emitted in more than 99.9\% of decays giving 2506~keV of deposited energy when both gammas get fully absorbed in the NC volume. Once the PMT parameters and the discriminator threshold were found with $^{60}$Co, a fine tuning of the gain was done with the central LED. Then, an AmBe source was used to verify the ability of the system to detect neutrons. An advantage of this neutron source is that in 60\% of the cases the ($\alpha,\!n$) reaction on Be leads to an excited $^{12}$C$^*$ state, which is then transfered to the ground state within 61~fs by emitting a gamma of 4438~keV. This gamma is a perfect trigger to start the acquisition window in order to look for a following neutron capture signal. In this respect it nicely resembles the primary trigger during a physics run and gives an additional reference for the energy calibration. It is also to be expected that some of the neutrons from AmBe source will first produce a prompt recoil proton and only then be captured on Gd. Thanks to the flash ADC used one can separately look into an energy spectrum of the first hit and any following one within the same event. As it can be seen from Fig.~\ref{fig:calib}, the primary hits form predominantly a bell-shaped spectrum specific to a prompt gamma while the secondary hits lead to a smoother spectrum which extends to higher energies, as one expects for partially escaping gammas appearing due to the neutron capture (see Sect.~\ref{sec:setup.all}). The same behaviour is also confirmed in GEANT4 simulations (toolkit version 9.2.p01) of AmBe source where elastic scattering off hydrogen, carbon and gadolinium nuclei, inelastic scattering on carbon, capture on gadolinium and hydrogen nuclei and an energy quenching present for less ionizing particles are taken into account (Fig.\ref{fig:calib}, \textit{crosses}). The initial kinetic energy of neutrons is generated according to an energy distribution of AmBe source. A simplified model with a common fixed threshold and an energy resolution scaled as $\Delta E(E)=\Delta E(4.4\:\mbox{MeV})\times\sqrt{E\:(\mbox{MeV})/4.4}\;$~(MeV) is used in the simulations, with a 2.8~MeV threshold and the energy resolution $\Delta E(4.4\:\mbox{MeV})=700$~keV as best fit. While the same values of the threshold and the resolution are used in the simulations to fit both measured ADC hits, the energy scaling of two spectra differs by 7\%. This is due to a position dependence of the light collection and the fact that 4.4~MeV $\gamma$'s are absorbed closer to the place of the AmBe source than neutron-capture gammas. Although the light propagation is not yet taken into account, a general good agreement between the measurement and the simulations is obtained (Fig.\ref{fig:calib}). Another verification of neutron capture on Gd is possible with the timing information, i.e. by looking into time intervals between the start of the event and any following hit (Fig.~\ref{fig:AmBe}). Neutrons of several MeV need first to thermalize in the medium of the detector and only then get captured following an exponential law with a characteristic time, the capture time, depending on the Gd content of the used scintillator. Measured data (Fig.~\ref{fig:AmBe}) confirm well this description. Fitting the falling part of the spectrum and considering a flat background, one extracts $\tau_{capture}=16.6\pm0.3\;\mu$s. The same time constant is obtained in GEANT4 (Tab.~\ref{tab:G4_Gd}) where various values of the Gd content can be considered. The simulated data are in good agreement with the measured $\tau_{capture}$ and 0.20\% of Gd admixture specified by the manufacturer (Tab.~\ref{tab:G4_Gd}). For further comparison we use the mean time of the measured distribution, $\left\langle t\right\rangle_{AmBe}=18.5\;\mu$s, because it accounts also for the effect of neutron moderation and can be easily used in case of low statistics.

\begin{table}
\caption{Capture time simulated in GEANT4 as a function of Gd content. In order to extract $\tau_{capture}$ the same interval and binning as in the measured data were used for fitting. No background is assumed.}
\label{tab:G4_Gd}       
\resizebox{0.49\textwidth}{!}{%
\begin{tabular}{lccccc}
\noalign{\medskip}
\hline\noalign{\smallskip}
Gd content, \% & 0.18 & 0.19 & 0.20 & 0.21 & 0.22 \\
\noalign{\smallskip}\hline\noalign{\smallskip}
$\tau_{capture}$, ns & 17800 & 17100 & 16290 & 15600 & 15000 \\
$\Delta\tau_{capture}$, ns & $\pm$100 & $\pm$100 & $\pm$90 & $\pm$80 & $\pm$80 \\
\noalign{\smallskip}\hline
\end{tabular}}
\end{table}

The calibration runs using the $^{60}$Co and AmBe sources are regularly executed in order to monitor the response of the detector under well-defined circumstances. 

\section{Preliminary physics data}
\label{sec:meas}

\begin{figure*}
    \begin{minipage}[c]{0.39\linewidth}
        \centering
        \resizebox{0.95\textwidth}{!}{%
           \includegraphics{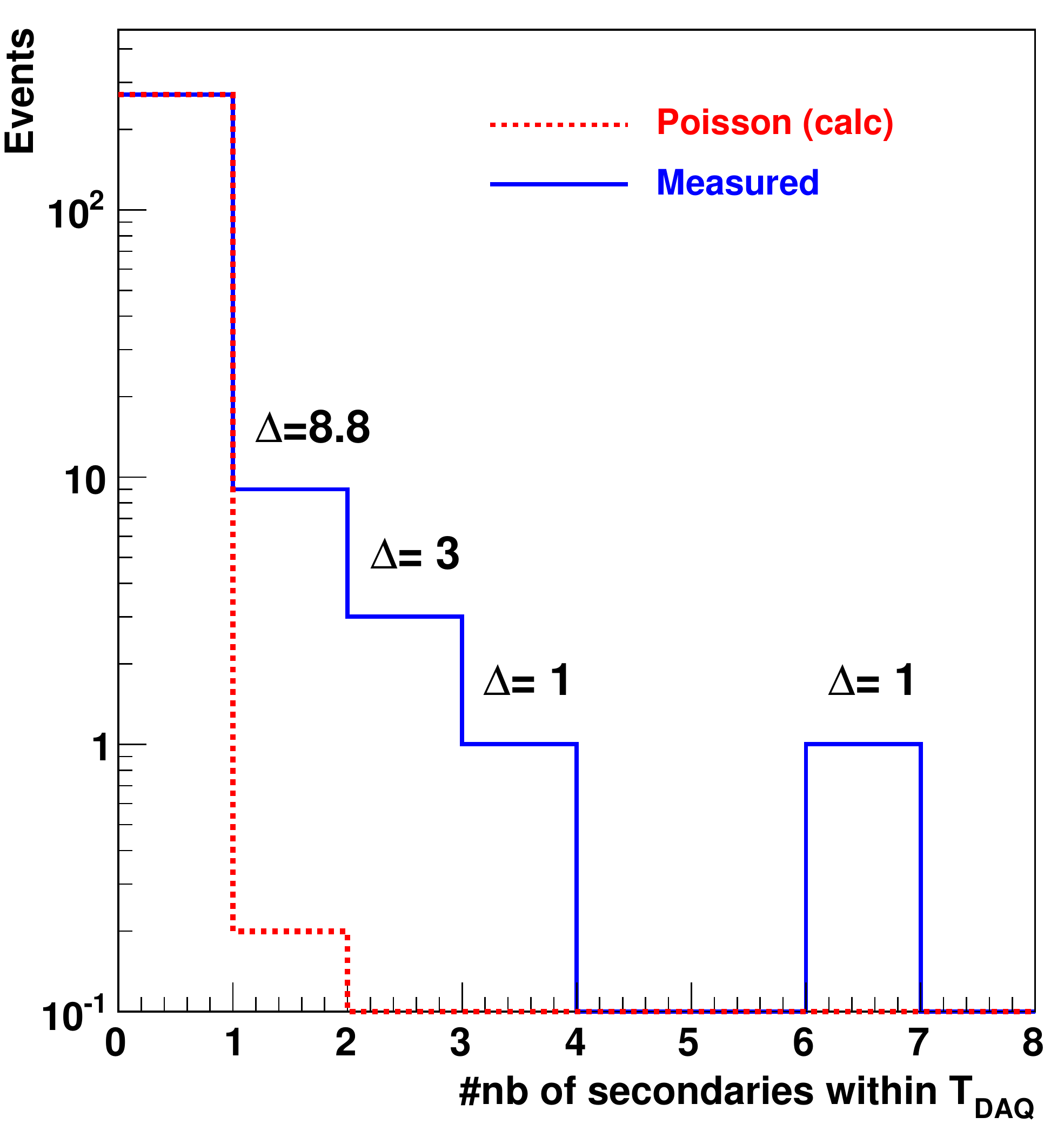}
        }
    \end{minipage}
    \begin{minipage}[c]{0.602\linewidth}
        \centering
        \resizebox{0.95\textwidth}{!}{%
           \includegraphics{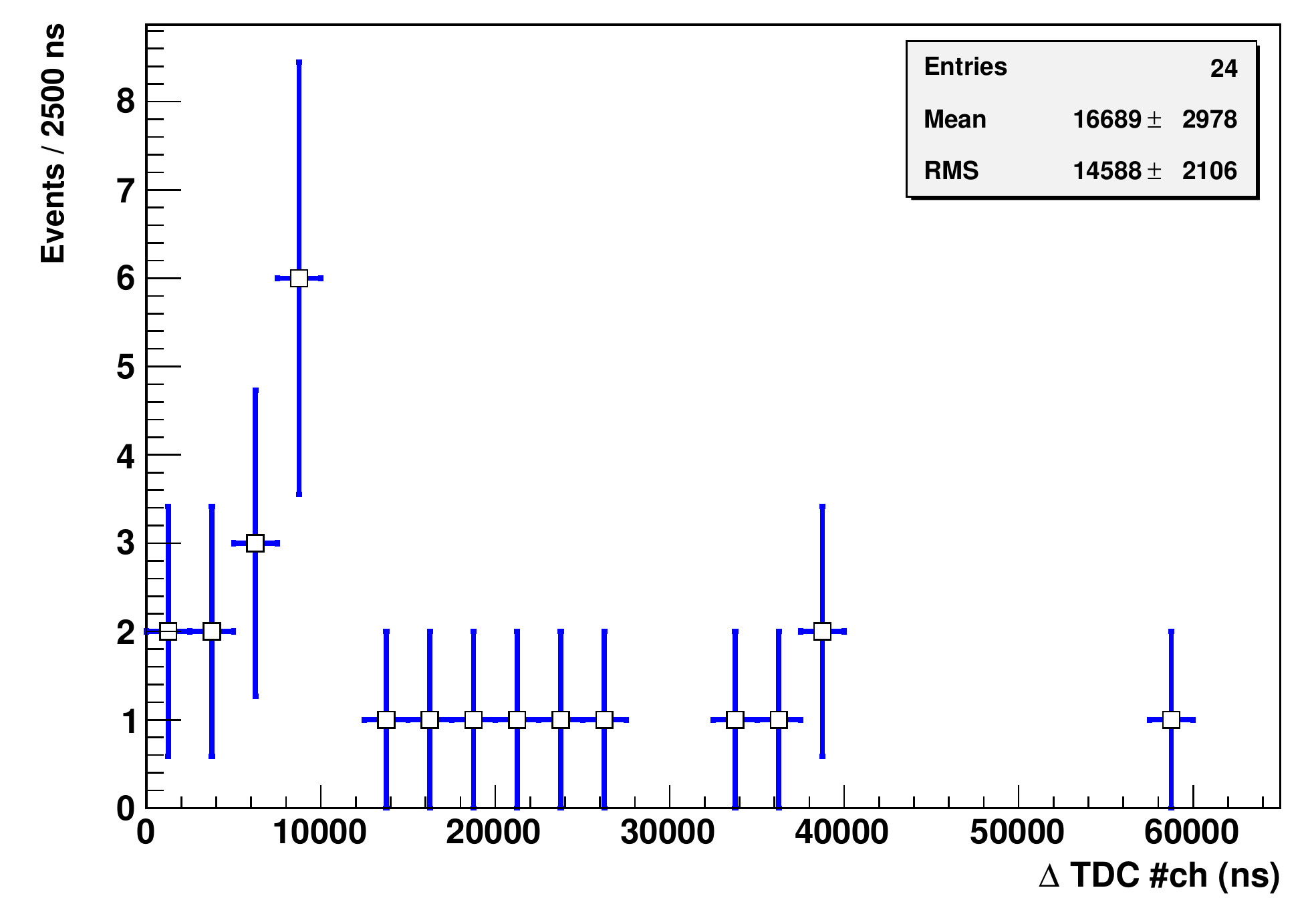}
        }
    \end{minipage} 
\caption{First data from a physics run, events selected with the $\mu$Veto module on top of the NC and the \textit{muon PMT's} first: number of secondaries within one event vs. Poisson expectation (\textit{left}) and the time intervals between first and any following TDC hit (\textit{right}). Live time of both systems running together is 51 days.}
\label{fig:phys4}  
\end{figure*}
\begin{figure*}
   \begin{minipage}[c]{0.39\linewidth}
        \centering
        \resizebox{0.95\textwidth}{!}{%
           \includegraphics{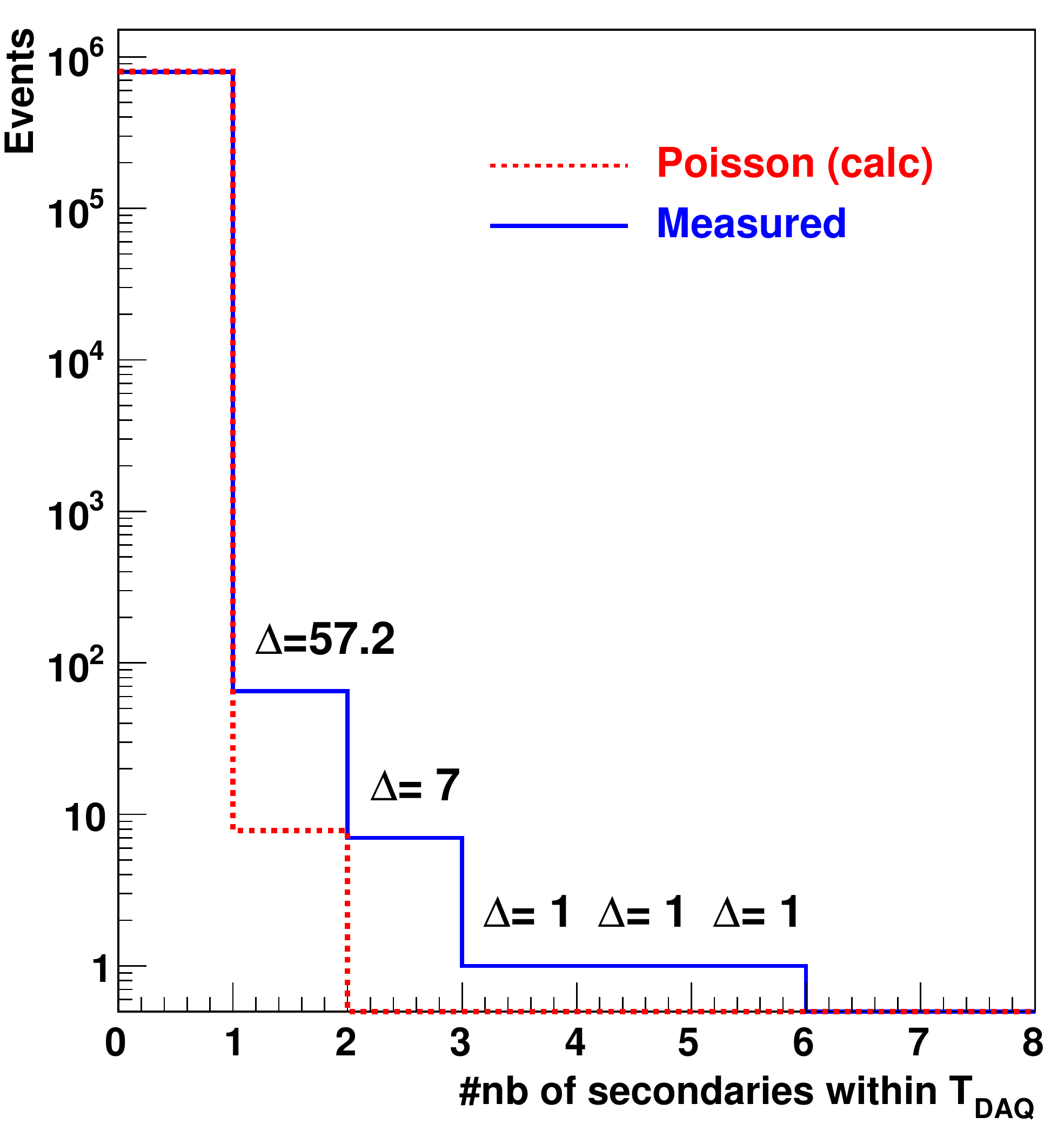}
        }
    \end{minipage}
    \begin{minipage}[c]{0.602\linewidth}
        \centering
        \resizebox{0.95\textwidth}{!}{%
           \includegraphics{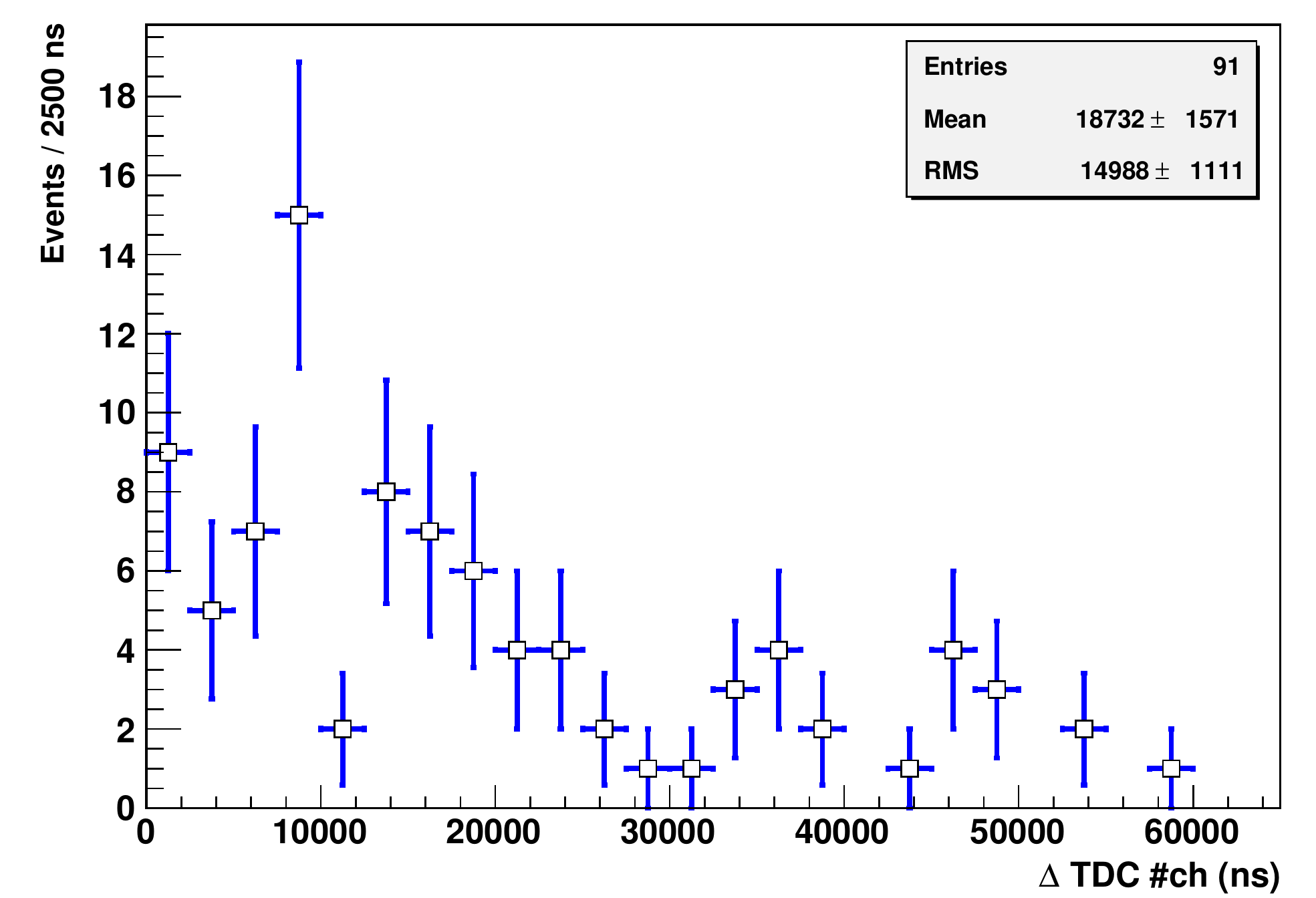}
        }
    \end{minipage} 
\caption{Preliminary data from a physics run, only neutron counter selected events: number of secondaries within one event vs. Poisson expectation (\textit{left}) and the time intervals between first and any following TDC hit (\textit{right}). Live time of the NC for this data set is 56 days.}
\label{fig:phys5}  
\end{figure*}
As a proof-of-principle we show in the following a preliminary analysis of physics runs. The muon-induced neutrons can be identified by the following features as described before: they produce multiple correlated hits within one event with distribution of time intervals similar to the one obtained for the AmBe source (Fig.~\ref{fig:AmBe}). The energy distribution of secondary ADC hits extends to higher values, i.e. well above the background spectrum. In the first analysis, we identify a prompt muon together with correlated secondary neutrons as to study the neutron production by the well-tagged muons. The muon candidates are defined by a coincidence of the $\mu$Veto module on top of the neutron detector and the \textit{muon PMT's}. Then, by using the time stamps of these events, a corresponding set of NC data is selected. Fig.~\ref{fig:phys4}(\textit{left}) shows the measured number of secondaries following a muon event and an expectation of Poisson distributed background of accidental coincidences based on the single hit rate. Only those secondaries are accepted which are at least 200~ns apart from the primary hit and each other, in order to avoid an effect of PMT afterpulses. As can be seen, there is a clear excess of measured secondaries above the background ($\Delta$-values on Fig.~\ref{fig:phys4}(\textit{left})). The corresponding time distribution of these secondary hits (Fig.~\ref{fig:phys4}(\textit{right})) has a mean value $\left\langle t\right\rangle=17\pm3\;\mu$s, which is clearly different from a central value of 29.8~$\mu$s specific for a random hit distribution within the DAQ window of 59.6~$\mu$s, and in agreement with $\left\langle t\right\rangle_{AmBe}=18.5\;\mu$s. Furthermore, we performed a Kolmogorov-Smirnov (KS) test on unbinned data of Fig.~\ref{fig:AmBe} and Fig.~\ref{fig:phys4}(\textit{right}). This yields \textit{prob.(KS)}=0.1 showing consistency of both spectra at a significance level of $\alpha$=0.05. In overall, within the 51 live days analysed here we identify 24 neutrons detected after the muons with an accidental background expectation of 0.2. This gives us a count rate of about 0.5 muon-induced neutrons per day.

To search for neutrons produced due to untagged muons, we look into neutron counter data only and multiple hit events. This is possible if the \textit{neutron PMT's} are tuned in such a way that the rate of accidental coincidences is well below an expected rate of muon-induced neutrons. To fulfill this requirement, half of the \textit{neutron PMT's} was set to a reduced gain. Fig.~\ref{fig:phys5} presents corresponding distributions of the number of secondaries and of related time intervals. As in the first analysis, the time cut of 200~ns is applied between consecutive hits.  As one can see, there is again a clear excess of measured data above the Poisson-based expectation of the background ($\Delta$-values on Fig.~\ref{fig:phys5}(\textit{left})). The mean value $\left\langle t\right\rangle=18.7\pm1.7\;\mu$s is also in good agreement with $\left\langle t\right\rangle_{AmBe}=18.5\;\mu$s. As above, KS-test is applied on unbinned data of Fig.~\ref{fig:AmBe} and Fig.~\ref{fig:phys5}(\textit{right}) and results in \textit{prob.(KS)}=0.3, denoting consistent spectral shapes at a significance level $\alpha$=0.05. The live time of this data set is 56~days. The data selection results in 75 events with minimum one secondary produced, among which there are 10 events with 2 or more secondaries detected. The expected accidental background is 7.8 and contributes to events with only one secondary particle. First simulations show that correlated background from ambient neutrons contributes less than 10 events and from the Bi-Po decay chain less than 1 event, both with 1 secondary only. The full sample was also checked for a coincidence with the $\mu$Veto module on top of the neutron detector and the \textit{muon PMT's}. Ten events with at least one secondary could be attributed to a tagged muon. Thus, the excess shown in Fig.~\ref{fig:phys5}(\textit{left}) clearly indicates muon-induced neutrons without requiring to detect the primary muon.

It is evident, that the first physics data show qualitatively the ability of the system to register muon-induced neutrons while a quantitative analysis of the production rate of these neutrons requires longer time of running to gather larger statistics and further detailed simulations to extract accurate detection efficiencies.

\section{Conclusion and outlook}
\label{sec:sum}
A neutron detection system based on 1~m$^3$ of Gd-loaded scintillator has been described. The purpose of this dedicated instrument is to measure specifically the muon-induced neutron background in the environment of present and planned dark matter experiments, e.g. EDELWEISS-II and EURECA, hosted by the LSM underground laboratory. This background is of concern for the DM searches and thus the neutron detector is \textit{in-location} of the DM setups. The NC system was installed in fall 2008 and, as the calibration and first physics runs show, it is fully operational. The aim of the measurement is to detect, in particular, the neutron yield in Pb both with or without tagging the primary muons. More detailed and accurate Monte-Carlo simulations taking into account all the aspects, e.g. the muon interaction, the neutron production, a light propagation, are under further development. The preliminary analysis of physics data show that the count rate of muon-induced neutrons in the order of 1~neutron/day has been achieved. Data-taking will continue through 2010 and early 2011 to get sufficiently large statistics.

\section*{Acknowledgments}
The help of the technical staff of the Karlsruhe Institute of Technology, N.~Bechtold, A.~Felden, S.~Jokisch and G.~Pro\-kott, and of the Laboratoire Souterrain de Modane is gratefully acknowledged. This work is in part supported by the German Research Foundation (DFG) through the Trans\-regional Collaborative Research Center SFB-TR27 as well as by the EU contract RII3-CT-2004-506222 and the Russian Foundation for Basic Research (grant No. 07-02-00355-a).





\begin{thebibliography}{00}
\footnotesize 

\bibitem{edw10} E. Armengaud et al., in preparation; V.Yu.~Kozlov for the EDELWEISS Collaboration, EPS-HEP09, Krakow, Poland, 2009, PoS(EPS-HEP 2009)~091 (http://pos.sissa.it); V.~Sanglard for the EDELWEISS Collaboration, TAUP09, Rome, Italy, 2009, arXiv:0912.1196v1 [astro-ph.CO].
\bibitem{edw09dm}E. Armengaud et al., Phys. Lett. B \textbf{687}, (2010) 294; arXiv:0912.0805v1 [astro-ph.CO].
\bibitem{kraus07}H. Kraus et al., Nucl. Phys. B (Proc. Suppl.) \textbf{173}, (2007) 168.
\bibitem{edw09id}A. Broniatowski et al., Phys. Lett. B \textbf{681}, (2009) 305 and references there in; arXiv:0905.0753v1 [astro-ph.IM].
\bibitem{chantelauze09}A.~Chantelauze, Investigation of the muon-induced background of the EDELWEISS-II experiment, Ph.D. thesis, Universit\'e Blaise Pascal, Clermont-Ferrand, 2009.
\bibitem{kudryavtsev03}V.A. Kudryavtsev et al., Nucl. Instrum. Meth. A \textbf{505}, (2003) 688.
\bibitem{araujo05}H. M. Ara\'ujo et al., Nucl. Instrum. Meth. A \textbf{545}, (2005) 398.
\bibitem{mei06}D.-M.~Mei and A.~Hime, Phys.~Rev. D \textbf{73}, (2006) 053004.
\bibitem{geant4}S.~Agostinelli et al., Nucl.~Instr.~Meth. A \textbf{506}, (2003) 250.
\bibitem{fluka}A. Fass\'o, A. Ferrari, P.R. Sala, in: A. Kling, F. Barao, M. Nakagawa, L. Tavora, P. Vaz (Eds.), Proceedings of the
Monte Carlo 2000 Conference (Lisbon, October 23–26, 2000), Springer, Berlin, 2001, p. 159;
A. Fass\'o, A. Ferrari, J. Ranft, P.R. Sala, in: A. Kling, F. Barao, M. Nakagawa, L. Tavora, P. Vaz (Eds.), Proceedings
of the Monte Carlo 2000 Conference (Lisbon, October 23–26, 2000), Springer, Berlin, 2001, p. 995.
\bibitem{gorshkov74}G.V. Gorshkov et al., Sov. J. Nucl. Phys., Vol.18, No.1 (1974) 57 and references there in.
\bibitem{bergamasco73}L. Bergamasco, S. Costa, P. Picchi, Il Nuovo Cimento 13A, (1973) 403.
\bibitem{wulandari04}H. Wulandari, et al., hep-ex/0401032.
\bibitem{araujo08}H. M. Ara\'ujo et al., Astropart. Phys. \textbf{29}, (2008) 471; arXiv:0805.3110v1 [hep-ex].
\bibitem{rozov10}S.~Rozov et al., submitted to J. Phys. G: Nucl. Phys; arXiv:1001.4383v1~[astro-ph.IM].
\bibitem{hennings07}R.~Hennings-Yeomans, D.S.~Akerib, Nucl.~Instr.~Meth. A \textbf{574}, (2007) 89.
\bibitem{nemo305}R. Arnold et al.,	Phys. Rev. Lett. \textbf{95}, (2005) 182302.
\bibitem{supernemo09}I.~Nasteva et al., HEP~2009 (EPS-HEP 2009), Cracow, Poland, 16-22 Jul 2009; arXiv:0909.3167v1~[hep-ex]. 
\bibitem{pdg08}C.~Amsler et al. (Particle Data Group), Phys. Lett. B \textbf{667}, (2008) 1.
\bibitem{horn07}M.~Horn, Simulations of the muon-induced neutron background of the EDELWEISS-II experiment for Dark Matter search, Ph.D. thesis, Universit\"at Karlsruhe (TH), 2007 and FZKA scientific report 7391 (http://bibliothek.fzk.de/zb/berichte/FZKA7391.pdf)
\bibitem{chantelauze07}A.~Chantelauze for the EDELWEISS Collaboration, SUSY2007 Karlsruhe, Germany, 2007;  arXiv:0710.5849v1~[astro-ph].
\bibitem{berger89}Ch.~Berger et al., Phys.~Rev.~D \textbf{40}, (1989) 2163.
\bibitem{lemrani06}R.~Lemrani et al., J. Phys. Conf. Ser. \textbf{39}, (2006) 145.
\bibitem{chazal98}V.~Chazal et al., Astropart. Phys. \textbf{9}, (1998) 163.
\bibitem{toi98}R.B. Firestone et al., Table of isotopes. 8th edition, New York: Wiley, 1998.
\bibitem{auger04}J.~Abraham et al., Nucl.~Instr. Meth. A \textbf{523}, (2004) 50. 
 \bibitem{kozlov08}V.Yu. Kozlov for the EDELWEISS Collaboration, IDM08 Stockholm, Sweden, 2008, PoS~(idm2008)~086 (http://pos.sissa.it);  arXiv:0902.4858v1~[astro-ph.IM].

\end{thebibliography}
\end{document}